\documentclass[%
 reprint,
 amsmath,amssymb,
 aps,
]{revtex4-1}

\usepackage{graphicx}% Include figure files
\usepackage{dcolumn}% Align table columns on decimal point
\usepackage{bm}% bold math
\usepackage{feynmp}

\begin{document}

%\preprint{APS/123-QED}

\title{New perspective on the Unruh effect}

\author{J. A. Rosabal}

\affiliation{Fields, Gravity and Strings @CTPU, Institute for Basic Science, 55, Expo-ro, Yuseong-gu, Daejeon, Korea, 34126.}

\date{\today}

\begin{abstract}
In this work,  based on the worldline path integral representation of the vacuum energy in spacetime with a Lorentzian metric,  we provide a new but complementary interpretation of the Unruh effect. We perform the quantization of the massless free scalar field in  Rindler space specifying initial and final conditions. After quantization, the final outcome for the vacuum energy is interpreted as world line path integrals. In this picture we find that the Unruh radiation is made of real particles as well as real antiparticles. The prediction regarding the  presence of antiparticles in the radiation might open new lines for experimental detection of the effect.  We present a thought experiment which offers a clear picture and supports the new interpretation.
\end{abstract}

\maketitle

\section{\label{sec:1}Introduction}

Since the seminal work \cite{Hawking:1974rv, Hawking:1974sw}, the physics community has been struggling to figure out the origin of the Hawking radiation and how to solve  the paradoxes that it generates. Several interpretations have emerged  over the last forty years \cite{Hartle:1976tp, Susskind:1993ws, Susskind:1994sm, Kabat:1995eq, Almheiri:2012rt},  but the problem remains unsolved. An easier but still conceptually rich setup is the so called Unruh effect \cite{Fulling:1972md, Davies:1974th, Unruh:1976db}. A noninertial observer in flat space, having  proper constant acceleration,  i.e.,  a  Rindler observer   \cite{Rindler:1966zz},  measures a vacuum energy given by the Planck thermal distribution with a temperature proportional to the acceleration $T=\frac{\hbar \text{a}}{2\pi \text{c} k}$. In what follows, we set $\hbar=\text{a}=\text{c}=k=1$,
\begin{equation}
E^{R}_{vac}\propto \int_{0}^{\infty} d\nu \frac{\nu}{\text{e}^{\frac{\nu}{T}}-1}. \nonumber
\end{equation}
In contrast, an observer with a constant velocity, a Minkowski or inertial observer,  measures a vanishing vacuum energy $E^{M}_{vac}=0$.

From the canonical quantization point of view,  it has been understood \cite{Fulling:1972md} that the Rindler observer experiences a different vacuum, in other words, different initial conditions compared to the  Minkowski observer. Thus she ``sees'' particles  (radiation). However, the controversy around where these particles are coming from keeps on,  and it has not been fully understood yet, see,  for instance  \cite{CruzyCruz:2016kmi} and references therein.  

In this work, we provide a new interpretation to the Unruh effect which is complementary to the one originally proposed by Unruh  \cite{Unruh:1976db}.  Instead of performing the ordinary canonical quantization  where the initial  value of the field operator and its conjugate momentum have to be specified, we quantize the massless  free scalar field in Rindler space imposing  initial and final conditions at the points were the acceleration is turned on and off. The advantage of this quantization  scheme is that by means of the propagators, expressed as worldline path integrals,  between the initial and final states in Rindler space,  we can trace the particles running in loops in spacetime with a Lorentzian metric, which are the ones that contribute to the vacuum energy.

\section{\label{sec:2}Motivation}

Let us start this section by collecting some features of quantum field theory (QFT). Throughout the paper, we will use these ingredients to uncover a different facet of the Unruh effect. 

\begin{figure}[hbt]
\centering
  \includegraphics[width=.25\textwidth]{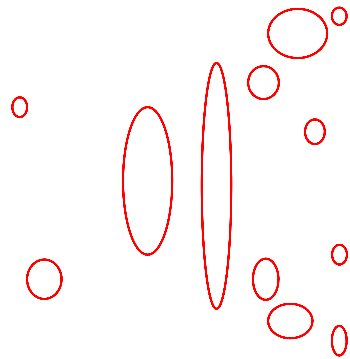}
  \caption{\sl Pictorial representation of the infinitely many  loops in spacetime contributing to the vacuum energy. No particular coordinate system is needed to describe them. }\label{fig1}
\end{figure}

In a noninteracting theory, the only processes that contribute to the vacuum energy are the loops in Fig. \ref{fig1}. These processes occur in spacetime, and they are events  that are independent of any observer (coordinate system).  
In the path integral formulation of QFT, the one loop partition function is indeed the vacuum energy of a given system. A clearer picture  comes out in the worldline path integral  representation of the vacuum energy where we can interpret the quantum particles  as moving over a collection of actual trajectories \cite{Feynman:1950ir,Strassler:1992zr,Daikouji:1995dz,Bastianelli:2002fv,  Witten:2015mec, E.Witten}.   For example,  the propagator can be represented as 
\begin{equation}
G(x_1,x_2)=\int_0^{\infty}ds\int_{x(0)=x_1}^{x(1)=x_2}Dx^{\mu}\text{exp}\Big[\text{i}\frac{1}{4s}\int_{0}^{1}d\tau\ \dot{\bold{x}}^2(\tau)\Big],  \nonumber
\end{equation}
where the functional integration is over all  paths starting at $x_1$ and ending at $x_2$ in Minkowski space and $\dot{\bold{x}}^2(\tau)=(\dot{x}^{0}(\tau))^2-(\dot{x}^{1}(\tau))^2-(\dot{x}^{2}(\tau))^2-(\dot{x}^{3}(\tau))^2$. The one loop partition function (vacuum energy) can also be written as a worldline functional  integral  as \cite{Daikouji:1995dz,Polchinski:1998rq}, 
\begin{equation}
E_{vac}\propto\frac{1}{2}\int_0^{\infty}\frac{ds}{s}\int_{PBC}Dx^{\mu}\text{exp}\Big[\text{i}\frac{1}{4s}\int_{0}^{1}d\tau\ \dot{\bold{x}}^2(\tau)\Big]\label{VE},
\end{equation}
where PBC stands for periodic boundary conditions.
Of course, these are formal expressions, but they allow us to think about the quantum processes in geometrical terms in Minkowski space, see Appendix A of \cite{Feynman:1950ir}. Although to make sense of them, a Wick rotation is needed.

A virtual particle in the Feynman approach of  QFT is associated with an internal propagator between two different points in  spacetime, or starting and ending at the same point (a loop).
The propagation of a  real particle, on the other hand,  is associated with a propagator (open paths), but, in addition,  a definite on shell external  momentum state is attached to it. 

For the sake of completeness, we present a brief description of Rindler  space \cite{Rindler:1966zz}. Rindler space  $M_R$ is defined as the region interior to the lines $x^0=x^1$  and $x^0=-x^1$,  with $x^1\geqslant0$,  in a two-dimensional Minkowski space. It can be extended to higher dimensions. In four dimensions, for instance,  the metric of $M_R$ can be written in two equivalent forms  
\begin{eqnarray}
ds^2 & = &   -\rho^2 d\tau^2+d\rho^2+(dx^2)^2+(dx^3)^2  \label{metric1}\\
 {} &  = &  \text{e}^{2 \xi}(-d\tau^2+d\xi^2)+(dx^2)^2+(dx^3)^2\label{metric2}.
\end{eqnarray}
These two metrics are related to the Minkowski one, $ds^2 = -(dx^0)^2+(dx^1)^2+(dx^2)^2+(dx^3)^2$, through the coordinate transformations 
\begin{eqnarray}\nonumber
x^0 (\rho,\tau) & = & \rho \ \text{sinh}(\tau)\\ \nonumber
x^1(\rho,\tau) & = & \rho \ \text{cosh}(\tau),
\end{eqnarray}
and
\begin{eqnarray}\nonumber
x^0(\xi,\tau) & = & \text{e}^{\xi} \ \text{sinh}(\tau)\\ \nonumber
x^1(\xi,\tau) & = & \text{e}^{\xi} \ \text{cosh}(\tau),
\end{eqnarray}
respectively.

\begin{figure}[hbt]
\centering
  \includegraphics[width=.3\textwidth]{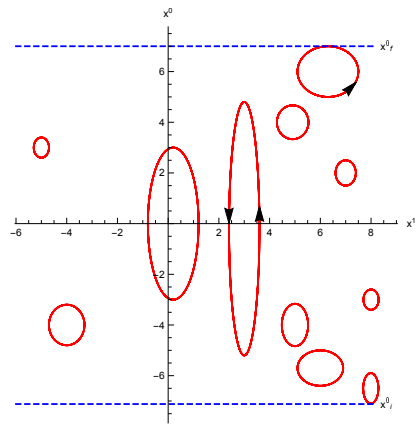}
  \caption{\sl Pictorial representation of the infinitely many  loops in spacetime relevant to the vacuum energy calculation  for an inertial observer. The blue lines are initial and final times in the Minkowski slicing. The arrows are representing the flow of time in the loop.}\label{fig2}
\end{figure}

Now let us perform a thought experiment in order to visualize how each observer perceives the vacuum processes and how they would compute the vacuum energy.
In the calculation of the vacuum energy, a Minkowski observer regards  all the loops in  spacetime as in Fig. \ref{fig2}.
 For a Rindler observer, on the other hand,  a naive calculation of the vacuum energy would proceed as follows. This observer would consider all the loops in $M_R$. However, this calculation would be incomplete since there are contributions that are not being taken into account. 

At this point, we stress that none of the observers actually see the virtual vacuum processes given by the particles running in loops.
Nevertheless, with the help of (\ref{VE}), they can build an intuitive and geometrical picture of these processes.

\begin{figure}[hbt]
\centering
  \includegraphics[width=.3\textwidth]{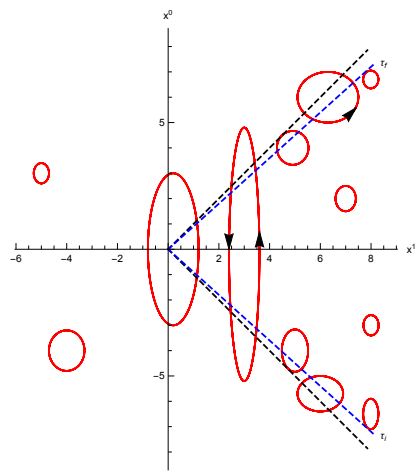}
  \caption{\sl Pictorial representation of the infinitely many  loops in spacetime as in Fig. \ref{fig2}, now in the context of the vacuum energy calculation for the  Rindler observer.}\label{fig3}
\end{figure}

\begin{figure}[hbt]
\centering
  \includegraphics[width=.18\textwidth]{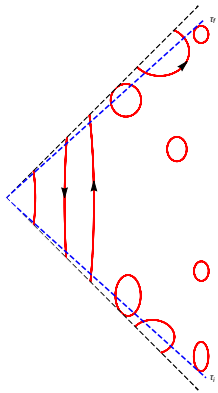}
  \caption{\sl Pictorial representation of the infinitely many  loops in spacetime. The Rindler observer perspective of the vacuum processes contributing to the vacuum energy. The arrows are representing the flow of time in the paths.}\label{fig4}
\end{figure}

To see the missing contributions, let us add to  Fig. \ref{fig2}  the Rindler horizons and an initial and final spacelike slice in $M_R$, as indicated in Fig. \ref{fig3}. The times  $\tau_i$ and $\tau_f$ can be placed anywhere in the Rindler wedge, even at the horizons,  $\tau_i\rightarrow -\infty$, $\tau_f\rightarrow \infty$. The portion of the spacetime the Rindler observer has access to is  limited by the lines $\tau_i$ and $\tau_f$, which mark the slices  were the acceleration is turned on and off.
Now to make it  easier to visualize the Rindler observer perspective,   let us detach  $M_R$ together with its content,  from the rest of the space, as in Fig. \ref{fig4}.
 From this we can see that the vacuum energy calculation from the Rindler observer point of view  has to take into account, in addition to all  loops in $M_R$,  the following contributions:
 \begin{enumerate}
 \item All the paths starting at   $\tau_i$  and ending at $\tau_i$, including the case where the initial and final  points on the slice $\tau_i$  are the same, i.e., the path is a loop.
 \item All the paths starting at   $\tau_f$  and ending at $\tau_f$, including the case where the initial and final  points on the slice $\tau_f$ are the same, i.e.,  the path is a loop.
  \item All the paths starting at   $\tau_i$  and ending at $\tau_f$ and vice versa.

 Notice that  from the Rindler perspective there will be paths running backward in time, representing real antiparticles.
 \end{enumerate} 

The picture described above suggests that while a Minkowski observer experiences a vacuum energy full of virtual particles (loops),  an observer in $M_R$ is able to measure a vacuum energy with real propagating particles. {\it From the non-inertial observer perspective, there are infinitely many paths that will never close into a loop;  thus, according to QFT, these paths will look to him as real particles. A more surprising consequence arising from this picture is that the Unruh radiation is made  of real particles as well as  real antiparticles. This fact may be inferred just by looking at the arrows in} Fig. \ref{fig4}.

Whether these contributions, Fig. \ref{fig4},  lead to a thermal radiation is something that we have to check. In the next section we will give a  mathematical proof of all the statements made in this section regarding the origin of the particles seen by the Rindler observer. 

\section{\label{sec:3} Quantization and Vacuum Energy}

In this section,  we compute the vacuum energy of a free scalar field in Rindler space. The usual canonical quantization in Minkoswki space is an initial value problem in which we solve the equation  $\Box \varphi=0$,   with the conditions  $\varphi(t_i)=\phi(t_i)$,  and $\partial_ t \varphi(t_i)=\psi(t_i)$. It is a well-posed Cauchy problem, and this data determines the function $\varphi$ for all $t$ in the region where we are solving the equation. In addition,  this initial condition is physical in the sense that we can always give the initial data at a given moment in time,  i.e.,  prepare  the initial state.  What would be non-physical is to give  initial  $\varphi(t_i)=\phi(t_i)$, and final $\varphi(t_f)=\phi(t_f)$,  conditions. While we can always prepare the initial state, the final state is impossible to prepare. Nevertheless, the solution of  $\Box \varphi=0$,  can be  fully determined  by the latter data too.  

In the Rindler space, the situation could be different. Since the Rindler patch does not cover the whole Minkowski space, the Rindler observer  might have access to the final data.   

Although the Rindler observer does not need to know explicitly the operator at the final time; she only needs to know the propagator in Rindler space, as we will see in the next section. The final state boundary condition for the Rindler observer can be motivated as follows. The Minkowski observer can  either solve the equation of motion $\Box \varphi_M(x^0,x^1,\bar{x})=0$   or evolve the initial field operator with his Hamiltonian,  
\begin{equation}
\varphi_M(x^0_f,x^1,\bar{x})=\text{e}^{\text{i}(x^0_f-x^0_i)H} \varphi_M(x^0_i,x^1,\bar{x})\text{e}^{-\text{i}(x^0_f-x^0_i)H}\ . \nonumber
\end{equation}
 This data can be communicated to the Rindler observer. 
 
 Starting with the action in the Rindler patch
\begin{multline}
S=\frac{1}{2}\int_{\tau_i}^{\tau_f} d\tau \int_{0}^{\infty}d\rho \int_{-\infty}^{\infty} d^2x\Big[ \\ 
\rho^{-1}(\partial_{\tau}\varphi_{R})^2-\rho\big((\partial_{\rho}\varphi_{R})^2+(\partial_{x^2}\varphi_{R})^2+(\partial_{x^3}\varphi_{R})^2 \big) \Big]\label{acctionRindler},
\end{multline}
she can quantize her system in two different and equivalent ways.  
 The ordinary one, where the field and its time derivative are specified at some initial time,  and she does not need the final data, 
\begin{enumerate}
\item 
\begin{eqnarray}\label{equM} \nonumber
\Box \varphi_R(\tau,\rho,\bar{x}) & = & 0 \\ \nonumber
  \varphi_R(\tau_i,\rho,\bar{x}) & = & \varphi_M(x^0(\tau_i,\rho),x^1(\tau_i,\rho),\bar{x})\\ \nonumber
  \partial_{\tau} \varphi_R(\tau_i,\rho,\bar{x}) & = & \partial_{\tau}\varphi_M(x^0(\tau_i,\rho),x^1(\tau_i,\rho),\bar{x}),
\end{eqnarray}
or
\item 
\begin{eqnarray}\label{equR}
\Box \varphi_R(\tau,\rho,\bar{x}) & = & 0 \\ \nonumber
  \varphi_R(\tau_i,\rho,\bar{x}) & = & \varphi_M(x^0(\tau_i,\rho),x^1(\tau_i,\rho),\bar{x})\\ \nonumber
  \varphi_R(\tau_f,\rho,\bar{x}) & = & \varphi_M(x^0(\tau_f,\rho),x^1(\tau_f,\rho),\bar{x}),
\end{eqnarray}
\end{enumerate}
where the field at the initial time and the field at the final time are specified.  A set of boundary conditions similar to
(\ref{equR}) was previously considered in \cite{Araya:2017evj}, but in a different context.

In both schemes,  we can impose the usual equal time canonical commutation relation,
\begin{eqnarray}\nonumber
 \big[\varphi_R(\tau,\rho,\bar{x}), \Pi_R(\tau,\rho,\bar{x}^{\prime})\big]   =
  \frac{\delta(\rho_i-\rho_f)}{\rho}\delta^2(\bar{x}-\bar{x}^{\prime}),
\end{eqnarray}
which holds for all times,  and $\Pi_R=\rho^{-1}\partial_{\tau}\varphi_{R}$. 

Here,  we choose to quantize the field using the  second option. In fact, the second option makes the variational problem of (\ref{acctionRindler}) inside the Rindler wedge a well-posed Lagrangian variational problem, where we fix the variations $\delta\varphi_{R}(\tau_i,\rho,\bar{x})=\delta\varphi_{R}(\tau_f,\rho,\bar{x})=0$ and as usual $\delta\varphi_{R}(\tau,\rho\rightarrow\infty,\bar{x}\rightarrow\pm\infty)=0$. The general  solution of $\Box \varphi_M=0$, in Minkowski space,  is 
\begin{multline}\label{modeexp}
\varphi_M(x^0,x^1,\bar{x})=\int_{-\infty}^{\infty}\frac{dk_1}{2\pi}\int_{-\infty}^{\infty}\frac{d^2k}{(2\pi)^2}\frac{1}{(2k_0)^{\frac{1}{2}}}\\
\Big(a_{(k_1,\bar{k})}\text{e}^{-\text{i}(k_0x^0+k_1x^1+\bar{k}\cdot\bar{x})}+a^{\dagger}_{(k_1,\bar{k})}\text{e}^{\text{i}(k_0x^0+k_1x^1+\bar{k}\cdot\bar{x})}\Big),
\end{multline}
while in the Rindler patch with the metric (\ref{metric1}), the general  solution of $\Box \varphi_R=0$  can be expressed  as
\begin{eqnarray}\nonumber
\varphi_R(\tau,\rho,\bar{x})=\int_{0}^{\infty}d\nu\int_{-\infty}^{\infty}\frac{d^2k}{(2\pi)^2}\frac{1}{(2\nu)^{\frac{1}{2}}}\\ \nonumber
\Big(b_{(\nu,-\bar{k})}\text{e}^{-\text{i}\nu \tau}+b^{\dagger}_{(\nu,\bar{k})}\text{e}^{\text{i}\nu \tau}\Big)\psi_{\nu,\bar{k}}(\rho)\text{e}^{\text{i}\bar{k}\cdot\bar{x}},
\end{eqnarray}
where $k_0=(k_1^2+|\bar{k}|^2)^{\frac{1}{2}}$,  and $\psi_{\nu,\bar{k}}(\rho)$,  are the normalized eigenfunctions of the equation 
\begin{equation}
\Big(\rho^2\frac{d^2}{d\rho^2}+\rho\frac{d}{d\rho}-|\bar{k}|^2 \rho^2+\nu^2\Big)\psi_{\nu,\bar{k}}(\rho)=0,\nonumber
\end{equation}
\begin{equation}\label{function}
\psi_{\nu,\bar{k}}(\rho)=\pi^{-1}\big(2\nu \text{sinh}(\pi \nu)\big)^{\frac{1}{2}}\text{K}_{\text{i}\nu}(|\bar{k}|\rho),
\end{equation} 
with $\text{K}_{\text{i}\nu}(|\bar{k}|\rho)$,  the modified Bessel function of the second kind \cite{Fulling:1972md}. By imposing (\ref{equR}) we get two equations of the form
\begin{eqnarray}\nonumber
\text{e}^{-\text{i}\nu \tau_{i,f}}b_{(\nu,-\bar{k})}+\text{e}^{\text{i}\nu \tau_{i,f}}b^{\dagger}_{(\nu,\bar{k})}= \\ \nonumber
(2\nu)^{\frac{1}{2}}\int_{0}^{\infty}d\rho\int_{-\infty}^{\infty}d^2x\frac{\psi_{\nu,\bar{k}}(\rho)}{\rho}\text{e}^{-\text{i}\bar{k}\cdot\bar{x}}\varphi_{M}(\tau_{i,f},\rho,\bar{x}),
\end{eqnarray}
where $\varphi_{M}(\tau_{i,f},\rho,\bar{x})$,  is the field operator as seen by the Minkowski observer but evaluated at the  constant $\tau=\tau_i,\tau_f$ slices.

Let us recall that the energy $E_{vac}^R$ seen by the Rindler observer  is given by
\begin{equation}
E_{vac}^R=\int_{-\infty}^{\infty} \frac{d^2k}{(2\pi)^2}\int_{0}^{\infty} d\nu\ \nu \  \langle 0 | b^{\dagger}_{(\nu,\bar{k})} b_{(\nu,\bar{k})}|0\rangle\ +E_{0}^R, \nonumber
\end{equation}
where  the term 
\begin{equation}
E_{0}^R = \frac{1}{2}\delta(0)^3 \int_{-\infty}^{\infty}d^2k  \int_{0}^{\infty}d\nu\nu,
\end{equation}
comes only from the loops inside the wedge. As in the Minkowski calculation, for most of the purposes, the term $E_{0}^R$ can be discarded. 
After solving for $b$ and $b^{\dagger}$ and plugging the solution into the number of particles $\bold{n}$, we get
\begin{eqnarray}\label{number}
\bold{n}=\langle 0 | b^{\dagger}_{(\nu,\bar{k})} b_{(\nu,\bar{k})}|0\rangle=\frac{\nu}{2 \big(\text{sin}[\nu(\tau_f-\tau_i)]\big)^2}\\ \nonumber
\times\int_{0}^{\infty}d\rho\int_{-\infty}^{\infty}d^2x\frac{\psi_{\nu,\bar{k}}(\rho)}{\rho}\text{e}^{-\text{i}\bar{k}\cdot\bar{x}}\ \ \ \\ \nonumber
\times\int_{0}^{\infty}d\rho^{\prime}\int_{-\infty}^{\infty}d^2x^{\prime}\frac{\psi_{\nu,\bar{k}}(\rho^{\prime})}{\rho^{\prime}}\text{e}^{\text{i}\bar{k}\cdot\bar{x}^{\prime}}\Big[ \\ \nonumber
\langle 0 |\varphi_{M}(\tau_i,\rho,\bar{x})\varphi_{M}(\tau_i\ ,\rho^{\prime},\bar{x}^{\prime})  |0\rangle \\ \nonumber
+\langle 0 |\varphi_{M}(\tau_f,\rho,\bar{x})\varphi_{M}(\tau_f,\rho^{\prime},\bar{x}^{\prime})  |0\rangle\\ \nonumber
-\text{e}^{-\text{i}\nu(\tau_f-\tau_i)}\langle 0 |\varphi_{M}(\tau_i,\rho,\bar{x})\varphi_{M}(\tau_f,\rho^{\prime},\bar{x}^{\prime})  |0\rangle\\ \nonumber
-\text{e}^{+\text{i}\nu(\tau_f-\tau_i)}\langle 0 |\varphi_{M}(\tau_f,\rho,\bar{x})\varphi_{M}(\tau_i,\rho^{\prime},\bar{x}^{\prime})  |0\rangle\\ \nonumber
\Big].
\end{eqnarray}

 Now, from (\ref{number}) it is clear that the processes contributing to the vacuum energy are those listed in the previous section. Notice that the propagators, which is the only information needed for knowing the energy,  are given in terms of the field operator in  Minkowski space, but we will regard them  from the perspective of the Rindler observer. The propagator in Rindler space can be represented  as a functional integral over paths in $M_R$ only,  as in Fig. \ref{fig4}. Notice  in (\ref{number}) there are definite momentum states attached to each propagator \cite{Daikouji:1995dz} . These states could be consider as on shell states.
 
 Expression (\ref{number}) answers the question of where the particles in the radiation  are coming from.
 {\it The particles have been there, in the spacetime, as vacuum fluctuations (loops) all the time. However, by accelerating  the  Rindler observer breaks some loops in such a way that the particles trajectory  looks to him as open path. In addition, there are on shell definite momentum states attached to the propagators,   thus real propagating particles and antiparticles} \cite{Daikouji:1995dz} .  
 
The appearance of $\tau_i$ and $\tau_f$ in (\ref{number}) may be disturbing.   However, as we will show in the next section, the vacuum energy is independent of $\tau_i$ and $\tau_f$ and coincides with the Planck thermal distribution.  

\section{\label{sec:4}Vacuum Energy Distribution}

We shall explicitly  compute the vacuum energy $E_{vac}^R$.  To this end, we will first compute the integral 
\begin{eqnarray}\label{integral}
I=\int_{0}^{\infty}d\rho\int_{-\infty}^{\infty}d^2x\frac{\psi_{\nu,\bar{k}}(\rho)}{\rho}\text{e}^{-\text{i}\bar{k}\cdot\bar{x}}\ \ \ \\ \nonumber
\times\int_{0}^{\infty}d\rho^{\prime}\int_{-\infty}^{\infty}d^2x^{\prime}\frac{\psi_{\nu,\bar{k}}(\rho^{\prime})}{\rho^{\prime}}\text{e}^{\text{i}\bar{k}\cdot\bar{x}^{\prime}} \\ \nonumber
\times\langle 0 |\varphi_{M}(\tau_i,\rho,\bar{x})\varphi_{M}(\tau_f\ ,\rho^{\prime},\bar{x}^{\prime})  |0\rangle. \nonumber
\end{eqnarray}
After plugging in,  (\ref{modeexp}) and (\ref{function}),  and taking  into account $a_{(p_1^{\prime},\bar{p}^{\prime})}|0\rangle=0$,  and
\begin{equation}
\big[ a_{(p_1,\bar{p})}\ , a^{\dagger}_{(p_1^{\prime},\bar{p}^{\prime})}\big]=(2\pi)^3\delta(p_1-p_1^{\prime})\delta^2(\bar{p}-\bar{p}^{\prime}),  \nonumber
\end{equation}
we get 
\begin{eqnarray}\nonumber
I=\frac{2\nu}{\pi}\text{sinh}(\pi\nu)\delta^2(0)\int_{-\infty}^{\infty}dp\frac{1}{\sqrt{p^2+|\bar{k}|^2}}\\ \nonumber
\times\int_0^{\infty}d \rho\frac{\text{K}_{\text{i}\nu}(|\bar{k}|\rho)}{\rho}\text{e}^{-\text{i}\rho\big(\sqrt{p^2+|\bar{k}|^2}\ \text{sinh}(\tau_i)+p \ \text{cosh}(\tau_i)\big)}\\ \nonumber
\times\int_0^{\infty}d \rho^{\prime}\frac{\text{K}_{\text{i}\nu}(|\bar{k}|\rho^{\prime})}{\rho^{\prime}}\text{e}^{\text{i}\rho^{\prime}\big(\sqrt{p^2+|\bar{k}|^2}\ \text{sinh}(\tau_f)+p \ \text{cosh}(\tau_f)\big)}.\nonumber
\end{eqnarray}
This integral can be further reduced by the change of variable $p=|\bar{k}|\text{sinh}(z)$,  to
\begin{eqnarray}\nonumber
I=\frac{2\nu}{\pi}\text{sinh}(\pi\nu)\delta^2(0)\int_{-\infty}^{\infty}dz\\ \nonumber
\times\int_0^{\infty}d \rho\frac{\text{K}_{\text{i}\nu}(|\bar{k}|\rho)}{\rho}\text{e}^{-\text{i}\rho|\bar{k}|\text{sinh}(z+\tau_i)}\\ \nonumber
\times\int_0^{\infty}d \rho^{\prime}\frac{\text{K}_{\text{i}\nu}(|\bar{k}|\rho^{\prime})}{\rho^{\prime}}\text{e}^{\text{i}\rho^{\prime}|\bar{k}|\text{sinh}(z+\tau_f)}. \nonumber
\end{eqnarray}
Interestingly enough $I$ does not depend on $|\bar{k}|$. For $|\bar{k}|=0$,  extra care is needed,   but it is not difficult to check, using the asymptotic  form of the modified Bessel function of the second kind, that the integration gives the same result and it is independent of $|\bar{k}|$ when   $|\bar{k}|\rightarrow0$. For the massive scalar field,  this precaution is not needed.

The Laplace transform like integrals  in $\rho$ and $\rho^{\prime}$  can be easily computed. After some algebra,  it reduces to 
\begin{eqnarray}\nonumber
I=\frac{\pi}{2\nu\text{sinh}(\pi\nu)}\delta^2(0)\ \ \ \ \ \ \  \ \ \ \ \ \ \ \ \ \ \ \ \ \ \ \ \ \\ \nonumber
\times\int_{-\infty}^{\infty}dz\Big(\text{e}^{\text{i}\nu(\tau_f-\tau_i)}(\text{i})^{-2\text{i}\nu}+\text{e}^{-\text{i}\nu(\tau_f-\tau_i)}(\text{i})^{2\text{i}\nu}\\ \nonumber
+\text{e}^{\text{i}\nu(2 z+\tau_i+\tau_f)}+\text{e}^{-\text{i}\nu(2 z+\tau_i+\tau_f)} \Big).
\end{eqnarray}
The last two term can be dropped since after $z$-integration, they contribute with a Dirac delta function $\delta(\nu)$,  and the integral $I$ is defined for all $\nu$ with $\nu\neq 0$. To make sense these expressions, we have to remove the zero mode $\nu= 0$. Picking the branches $(\text{i})^{-2\text{i}\nu}=(\text{i})^{2\text{i}\nu}=\text{e}^{-\pi \nu}$,  we obtain
\begin{equation}
I=\frac{2\pi}{\nu}\frac{1}{\text{e}^{2\pi\nu}-1}\text{cos}\big[\nu(\tau_f-\tau_i)\big]\delta^2(0)\int_{-\infty}^{\infty}dz. \label{int}
\end{equation}
Plugging (\ref{int}) into (\ref{number}),  the time dependence cancels out, giving the desired result
\begin{equation}
\bold{n}=\frac{1}{2\pi}\frac{1}{\text{e}^{2\pi\nu}-1} V_3, \nonumber
\end{equation}
where $V_3$, is the volume of space and we have used  $\delta(0)=\frac{1}{2\pi}\int_{-\infty}^{\infty}dx$.

\section{\label{sec:5}Conclusions}

The worldline path integral representation of the vacuum energy provides an intuitive and visual framework for describing quantum processes. It makes it easier to translate these processes into geometrical terms. As in the thought experiment of the ray light in Einstein's train or elevator, we have performed a similar thought experiment which allowed us to visualize the reality that  a Rindler observer experiences. The main difference with the classical experiment of  Einstein is that we considered the quantum nature of the particles. In other words, we have taken into account all the possible paths in the spacetime associated with the vacuum energy. Putting all the ingredients together, the Unruh effect naturally  emerges. 

We have used the worldline path integral representation of the vacuum energy as motivation but the calculations  were done within the operator formalism of QFT and then reinterpreted as  worldline path integrals \cite{Daikouji:1995dz}.  We have quantized the free massless scalar field in a rather arguable,  but mathematically  well justified, fashion. Each step of the calculation admits a nice physical interpretation consistent with the thought experiment. Furthermore,  the final result is the one expected for the Unruh effect.   

It would be worthwhile to conduct the same calculation entirely within the worldline formalism in spacetime with Lorentzian metric. In this case, the vacuum energy would be given by a path integral where the integration is over two topologically different kinds of paths, i.e., those paths listed in section \ref{sec:2}  where the loops are restricted to the Rindler wedge.

If we believe in the premises of QFT, an irrefutable physical interpretation comes out from the picture presented above. {\it The particles that the Rindler observer experiences are those virtual particles moving in loops in the whole (Minkowski) space that from the Rindler observer's perspective never complete a loop, being seen as real particles and real antiparticles. We would like to emphasize  that this observation (the presence of antiparticles in the radiation) might open  new possibilities for experimental confirmation of the effect}.

We started this paper by citing Hawking's work on black holes, but we studied a flat space phenomenon called the Unruh effect that has many features in common.  The natural extension of this work should be including gravity into the game \cite{rosabal_wip}. In particular the black hole geometry, in the same spirit of \cite{Susskind:1993ws, Susskind:1994sm, Kabat:1995eq}. We believe that the physical interpretation presented here could shed some light onto the quantum aspects of  the black hole geometry and  gravity in general. Let us remark that the interpretation of this picture in the black hole geometry  would differ from that popular and conceptually inaccurate notion (see, for instance, the comments in page $31$ of  \cite{Harlow:2014yka})  where the antiparticle falls into the black hole and the particle runs away to infinity, draining mass from the hole.

\section*{Acknowledgments}

I am grateful  to the postdoc members of the ``Fields, Gravity and Strings''  group  for discussions. Especially to Pablo Diaz and  Vladislav Vaganov for discussions and useful  comments. I would also to thank  H. Casini for the reading of the manuscript and useful remarks.

\appendix
\section{\label{sec:6}Vacuum Energy Distribution for $D=2$ }

In two dimensions, it is convenient to use the two dimensional version of  the metric (\ref{metric2}). Because of conformal symmetry, the solution in both patches takes a similar form
\begin{multline}\nonumber
\varphi_M(x^0,x^1)=\int_{-\infty}^{\infty}\frac{dk}{\pi(8|k|)^{\frac{1}{2}}}\\ \nonumber
\Big(a_{k}\text{e}^{-\text{i}(|k|x^0+kx^1)}+a^{\dagger}_{k}\text{e}^{\text{i}(|k|x^0+kx^1)}\Big),
\end{multline}
and
\begin{equation}\nonumber
\varphi_R(\tau,\xi)=\int_{-\infty}^{\infty}\frac{dp}{\pi(8|p|)^{\frac{1}{2}}}\nonumber
\Big(b_{-p}\text{e}^{-\text{i}|p|\tau}+b^{\dagger}_{p}\text{e}^{\text{i}|p|\tau}\Big)\text{e}^{\text{i}p\xi}.
\end{equation}
In this case, the analogue of the integral (\ref{integral}) is
\begin{align}\nonumber
I & =  \frac{1}{4\pi}\int_{-\infty}^{\infty}\frac{dk}{|k|}\\ \nonumber
{} & \times \int_{-\infty}^{\infty}d\xi\text{exp}\big(-\text{i}\text{e}^{\xi}\big(|k| \text{sinh}(\tau_i)+k \ \text{cosh}(\tau_i)\big)-\text{i}p\xi\big)\\ \nonumber
{} & \times  \int_{-\infty}^{\infty}d\xi^{\prime}\text{exp}\big(\text{i}\text{e}^{\xi^{\prime}}\big(|k| \text{sinh}(\tau_f)+k \ \text{cosh}(\tau_f)\big)+\text{i}p\xi^{\prime}\big).
\end{align}
It can be easily solved,  giving
\begin{eqnarray}\nonumber
I & = & \frac{1}{4\pi}(-1)^{-\text{i}p}\Gamma(\text{i}p)\Gamma(-\text{i}p)\\ \nonumber
{} & {} & \int_{-\infty}^{\infty}\frac{dk}{|k|}\Big(\frac{|k| \text{sinh}(\tau_i)+k \ \text{cosh}(\tau_i)}{|k| \text{sinh}(\tau_f)+k \ \text{cosh}(\tau_f)}\Big)^{\text{i}p}.
\end{eqnarray}
Using $\Gamma(\text{i}p)\Gamma(-\text{i}p)=\frac{\pi}{|p|\text{sinh}(\pi p)}$, after some algebra, it reduces to 
\begin{equation}
I=\frac{1}{|p|}\frac{1}{\text{e}^{2\pi p}-1}\text{cos}\big[|p|(\tau_f-\tau_i)\big]\int_{0}^{\infty}\frac{dk}{k}\label{I2}.
\end{equation}
Plugging (\ref{I2}) into the number of particles (\ref{number}), in this case with $\nu\rightarrow |p|$,  we get
\begin{equation}
\langle 0 | b^{\dagger}_{p} b_{p}|0\rangle=\frac{1}{\text{e}^{2\pi |p|}-1}V_1,\nonumber
\end{equation}
where $V_1=\int_{0}^{\infty}\frac{dk}{k}=\int_{-\infty}^{\infty}dx$, is  the volume of space. It is exactly the result,  computed within the ordinary canonical quantization framework,  reported in the Appendix A of \cite{Lee:1985rp}.

\end{document}